\documentclass[lettersize,journal]{IEEEtran}
\usepackage{amsmath,amsfonts}
\usepackage{algorithmic}
\usepackage[linesnumbered,ruled]{algorithm2e}
\SetKwInOut{Input}{Input}
\SetKwInOut{Output}{Output}
\usepackage{array}
\usepackage[caption=false,font=normalsize,labelfont=sf,textfont=sf]{subfig}
\usepackage{textcomp}
\usepackage{stfloats}
\usepackage{url}
\usepackage{verbatim}
\usepackage{graphicx}
\usepackage{cite}
\usepackage{xcolor}
\begin{document}

\title{Ring Artifacts Removal Based on Implicit Neural Representation of Sinogram Data}

\author{Ligen Shi, Xu Jiang, Yunze Liu, Chang Liu, Ping Yang, Shifeng Guo, and Xing Zhao*
\thanks{Ligen Shi, Xu Jiang and Yunze Liu are with the School of Mathematical Sciences, Capital Normal University, Beijing 100048, China (e-mail: ligenshi0826@gmail.com; chianghsu97@gmail.com; s-lyz25@bjzgca.edu.cn).}
\thanks{Ping Yang is with the Institute of Nuclear and New Energy Technology, Tsinghua University, Beijing 100084, China (e-mail: yang\_ping0603@163.com).}
\thanks{Chang Liu is with the Institute of Applied Mathematics, Beijing Information Science and Technology University, Beijing 100101, China (e-mail: changliuct@gmail.com).}
\thanks{Shifeng Guo is with the Shenzhen Institutes of Advanced Technology, Chinese Academy of Sciences, Shenzhen 518055, China (e-mail: sf.guo@siat.ac.cn).}
\thanks{Xing Zhao is with the School of Mathematical Sciences, Capital Normal University, Beijing, 100048, China (e-mail: zhaoxing\_1999@126.com).}
\thanks{Corresponding author: Xing Zhao}}

\markboth{Journal of \LaTeX\ Class Files,~Vol.~, No.~, ~2024}%
{Ligen Shi \MakeLowercase{\textit{et al.}}: Ring Artifacts Removal Based on Implicit Neural Representation of Sinogram Data}

\IEEEpubid{0000--0000/00\$00.00~\copyright~2024 IEEE}

\maketitle

\begin{abstract}
Inconsistent responses of X-ray detector elements lead to stripe artifacts within the sinogram data, which subsequently manifest as ring artifacts in the reconstructed computed tomography (CT) images, severely degrading image quality. This paper presents a novel method for correcting stripe artifacts in the sinogram data by separating the sinogram into an Ideal Sinogram (IS) and Stripe Artifacts (SA), with both components parameterized through Implicit Neural Representations (INR). The proposed method leverages INR to correct defective pixel response values using implicit continuous functions while simultaneously learning stripe features in the angular direction of the sinogram data. These two components, IS and SA, are combined within an optimization constraint framework, achieving unsupervised iterative correction of stripe artifacts in the projection domain. Experimental results demonstrate that the proposed method significantly outperforms current state-of-the-art techniques in effectively removing ring artifacts while maintaining the clarity and fidelity of CT images, thereby enhancing the overall diagnostic quality of CT imaging.
\end{abstract}


\section{Introduction}
\label{sec:introduction}
\IEEEPARstart{R}{ing} artifacts are common artifacts in computed tomography (CT) imaging, primarily caused by inconsistent responses of detector elements to X-rays or damage to these elements \cite{hsieh2003computed, boas2012ct}. Photon Counting Detectors (PCDs)\cite{danielsson2021photon, nakamura2023introduction}, due to their photon energy discrimination capability, have the potential advantage of providing more imaging information compared to conventional CT detectors. However, their broader application is hindered by ring artifacts, as the detector elements are prone to damage\cite{persson2012framework}. Inconsistent detector element responses lead to data deviations at different view angles, resulting in stripes in the sinogram. During CT image reconstruction, these stripes back-project as equidistant lines from the rotation center and, when accumulated from multiple view angles, form ring structures, known as ring artifacts. Removing these ring artifacts, or ring artifact correction, is a key technique for improving the quality of CT imaging.

\IEEEpubidadjcol

\begin{figure}[!ht] 
\centering \includegraphics[width=0.48\textwidth]{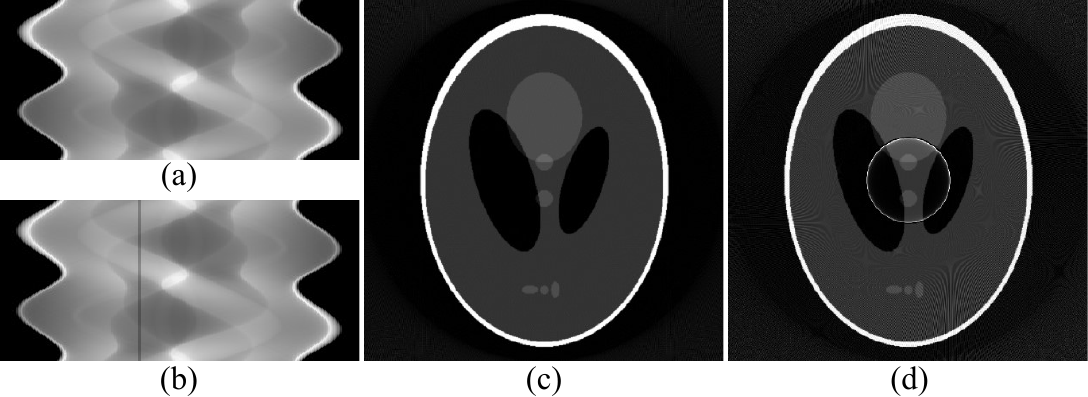} \caption{(a) Sinogram data without stripes; (b) Sinogram data with one stripe artifact; (c) and (d) are the FBP-reconstructed CT images from (a) and (b), respectively. The display range is [0, 1] for the sinogram and [0, 0.015] for the CT images.} \label{fig_0} 
\end{figure}

Widely adopted methods for ring artifact removal include detector calibration methods\cite{van2015dynamic, lifton2019ring, croton2019ring}, sinogram domain preprocessing methods\cite{Boin06, kowalski1978suppression, ashrafuzzaman2011self, raven1998numerical, munch2009stripe, makinen2022ring, Vo:18}, CT image post-processing methods\cite{sijbers2004reduction, chen2009ring, wei2013ring, huo2016removing, paleo2015ring}, dual-domain iterative methods \cite{salehjahromi2019new, zhu2024dual}, and data-driven methods \cite{chang2020cnn, liu2023detector, wang2019removing, zhao2018removing}. Detector calibration includes dark-field correction, flat-field correction, and bad-pixel correction. However, equipment aging can lead to deviations in correction coefficients, making it difficult to completely remove ring artifacts by relying solely on detector calibration. To effectively remove these artifacts, detector calibration methods need to be combined with other methods to achieve better correction results. CT image post-processing methods involve Cartesian and polar coordinate transformations, potentially leading to a loss of image resolution. Dual-domain iterative methods leverage features from both the projection domain and the CT image domain to establish optimization functions through iterative solving. However, since the stripe artifacts in the sinogram represent high-frequency information, they manifest not only as ring artifacts in the reconstructed images but also spread throughout the image space, causing additional noise and artifacts. As shown in Fig.~\ref{fig_0}, (a) shows sinogram data without stripes, and (b) shows sinogram data with a single stripe. (c) and (d) show the Filtered back-projection (FBP)\cite{herman2009fundamentals}  reconstructions of (a) and (b), respectively. Comparing Figs.~\ref{fig_0} (c) and (d), besides the ring artifacts, Fig.~\ref{fig_0} (d) also shows other noticeable artifacts. Therefore, it is challenging to achieve satisfactory results using single regularization constraints in the image domain. Data-driven supervised learning methods for artifact removal heavily depend on the size and quality of labeled datasets.

For sinogram-domain preprocessing methods, they indirectly eliminate ring artifacts in the reconstructed image by removing stripe artifacts in the sinogram. These methods directly leverage artifact features in the sinogram, avoiding potential information loss introduced during CT image reconstruction. However, existing sinogram-domain methods still face several challenges in stripe artifact removal, including: 1) loss of useful high-frequency information in the sinogram, resulting in blurry reconstructed CT images; 2) inaccuracies introduced when handling stripes caused by non-responsive detector elements with local interpolation; and 3) the lack of high-quality datasets, limiting the application of supervised learning methods.

To address these challenges, and given that sinogram data is continuous\cite{wu2013continuity}, this paper employs Implicit Neural Representation (INR) for its parametric representation. INR \cite{2020NeRF, SIREN, NeFieldsinVisualComputingBeyond, essakine2024we, liu2024finer, kazerouni2024incode, saragadam2023wire, sun2021coil} models data as a continuous implicit function, offering advantages such as resolution invariance, memory efficiency, and generalization beyond discrete data structures. CoIL\cite{sun2021coil} introduced the use of INR for sinogram restoration. Similarly, we adopt INR to modify the unresponsive pixels in the sinogram and develop a method for ring artifact removal. The proposed method effectively removes ring artifacts while preserving the details of reconstructed images, without the need for labeled datasets. 

\begin{figure}[!ht] 
\centering 
\includegraphics[width=0.48\textwidth]{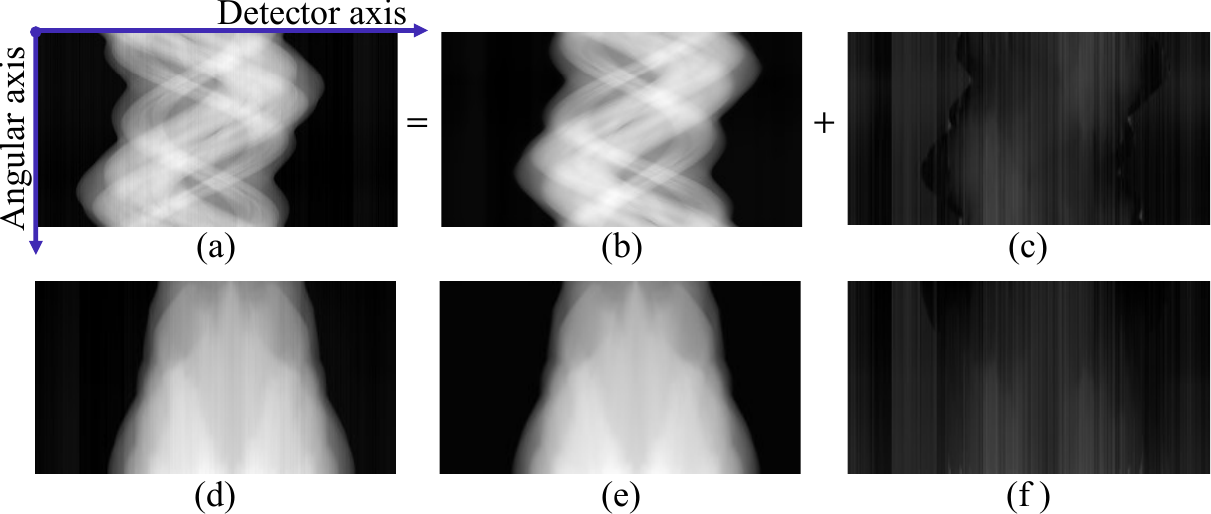} 
\caption{(a) Sinogram data with stripe artifacts; (b) IS; (c) SA. (d), (e), and (f) are the sorted results of (a), (b), and (c) along the angular axis based on projection values. The display range is [0, 3.07] for (a), (b), (d), and (e), and [-0.5, 0.5] for (c) and (f). The Angular axis represents the angular direction, and the Detector axis represents the detector direction.} \label{fig_1} \end{figure}

The response of X-ray detector elements with inconsistent responses can be assumed as\cite{salehjahromi2019new}:
\begin{align}\label{Eq1_}
I_{m}=C_m \cdot I_0 \cdot \mathrm{e}^{-\int_{L} \mu(\mathbf{x}) \mathrm{d}l},
\end{align}
where $C_m$ represents the gain factor between the measured count of the $m$-th detector element and the incident photon count, which is related to the response function of the detector element. $I_0$ and $I_m$ represent the number of emitted and received photons, respectively. Eq.~\eqref{Eq1_} can be rewritten as:
\begin{align}
-\log(I_m/I_0) = \int_{L} \mu(\mathbf{x}) \mathrm{d}l + \log(1/C_m),
\end{align}
where $\int_{L} \mu(\mathbf{x}) \mathrm{d}l$ represents the ideal projection, and $\log(1/C_m)$ corresponds to the stripe artifacts for the $m$-th detector element. Thus, the sinogram combines the Ideal Sinogram (IS) and Stripe Artifacts (SA), as shown in Fig.~\ref{fig_1}. 

Digital Radiography (DR) is primarily affected by scattering during the imaging process, resulting in low contrast while exhibiting local smoothness characteristics \cite{DR_OE_2015}. These properties are well-suited for parameterization by neural networks \cite{sun2021coil, zang2021intratomo}. The inconsistent responses of detector elements to X-rays lead to substantial variations at the same scan view angle, resulting in noticeable stripes in the sinogram; nevertheless, local smoothness in IS is maintained in the detector direction. While the detector responses vary, the intensity fluctuations of X-rays received by the same detector element at adjacent view angles are minimal, resulting in sparse gradients in the angular direction. Inspired by the sorted sinogram method proposed in \cite{Vo:18}, the sorting of Figs.~\ref{fig_1} (a), (b), and (c) along the angular axis by the projection values results in Figs.~\ref{fig_1} (d), (e), and (f). Sorting significantly enhances the smoothness of IS in the detector direction and further emphasizes the sparsity of SA in the angular direction.

Based on this analysis, the sinogram is modeled as a combination of IS and SA, constructing a differentiable forward problem model. By incorporating the local smoothness of IS in the detector direction and the sparse distribution of SA in the angular direction, we propose a unidirectional regularization framework that separates IS and SA.

The proposed method first classifies the pixels in the projection data into defective pixels (caused by non-responsive detector elements) and non-defective pixels (resulting from inconsistent X-ray detector responses). For non-defective pixels, the projection data is decomposed into the IS and SA under the unidirectional regularization framework, achieving the parameterization of IS from a ``discrete-to-continuous" representation. For defective pixels, the implicit function fitted to the INR-parameterized IS is used for prediction, thereby avoiding inaccuracies introduced by local interpolation. Additionally, to avoid the loss of useful high-frequency information in the sinogram, a residual compensation strategy is employed to further enhance image details. The main contributions of this paper are as follows: 
\begin{enumerate} 
\item We propose an unsupervised method for stripe artifact removal, which eliminates stripe artifacts in the sinogram domain without relying on labeled datasets. 
\item We leverage the continuous representation capability of INR-parameterized continuous functions to fit IS using data from non-defective pixels, enabling the prediction of projection values for defective pixels.
\item We introduce a unidirectional regularization framework for the effective separation of IS from SA. The regularizer $\Psi_{IS}$ enforces local smoothness within IS, while $\Psi_{SA}$ accentuates stripe features in SA and suppresses non-stripe artifacts, resulting in efficient separation of these two components.
\end{enumerate}

\section{Related Work}
In the sinogram domain, methods for removing stripe artifacts can be further categorized into frequency and spatial domain filtering. Stripe artifacts manifest as high-frequency components in the frequency domain of the sinogram, making frequency domain filtering a commonly used method for their removal. For example, high-pass filtering\cite{kowalski1978suppression} is used to correct projections by filtering out high-frequency components in the sinogram. Additionally, methods such as the variable window moving average and weighted moving average filters\cite{ashrafuzzaman2011self}, as well as Butterworth low-pass filtering\cite{raven1998numerical}, are also employed. The combined wavelet-FFT filtering method\cite{munch2009stripe} compresses the stripe information into narrow bands and eliminates it by multiplying with a Gaussian damping function. While these methods can remove stripe artifacts to some extent, they often result in information loss in the projection, thereby reducing the quality of the reconstructed images. Spatial domain filtering methods mainly use edge-preserving smoothing filters on the sinogram. The sorting-based method\cite{Vo:18} removes stripe artifacts by applying median filtering to the sorted sinogram, while the combination-based method\cite{Vo:18} identifies and interpolates the defective pixels through linear fitting, followed by filtering. For removing ring artifacts in PCD, Wu et al.\cite{an2020ring} generated a mean vector by averaging multi-angle projection images, applied a TV-$\ell_1$ aG filter to smooth the mean vector, and compensated the filtered mean vector back into the sinogram to eliminate stripe artifacts. However, the effectiveness of these methods depends on the accurate localization and compensation of stripe artifacts.

In CT image post-processing, a common technique involves converting images from Cartesian to polar coordinates. This transformation turns ring artifacts into stripe-like artifacts that are easier to identify and address. After processing in polar coordinates, the image is converted back to Cartesian coordinates to remove the ring artifacts. For example, Sijbers et al.\cite{sijbers2004reduction} use morphological operators to extract regions of interest (ROIs) with ring artifacts, convert them to polar coordinates, and identify homogeneous rows using a sliding window and signal variance. They then create artifact templates and correct them by subtracting them from each row of the polar image, followed by conversion back to Cartesian coordinates. Chen et al.\cite{chen2009ring} apply Independent Component Analysis (ICA) in the polar coordinate system to decompose CT images, use smoothing filters to address noise components, and reconstruct the image by combining ICA components before transforming it back to Cartesian coordinates. Liang et al.\cite{liang2017iterative} propose an iterative method that generates residual images by subtracting smoothed images, filtered with Relative Total Variation (RTV), from the originals in polar coordinates, then removes average artifacts and merges the details back into the original image. Wu et al.\cite{wu2019removing} develop an iterative ring artifact removal method using Unidirectional Total Variation (UTV) and TV-Stokes filtering. Paleo et al.\cite{paleo2015ring} present methods within the Total Variation and Dictionary Learning frameworks. However, filtering and correction based on polar coordinates may reduce image resolution due to coordinate transformations and interpolation.

Dual-domain iterative methods use sparse constraints on stripe artifacts in the sinogram and smoothness constraints on CT images, combining various regularization terms to establish an optimization function for iterative solving. Yu et al.\cite{salehjahromi2019new} introduced a ring total variation regularization term to the optimization model to penalize ring artifacts in the image domain. A correcting vector was also developed to address detector malfunctions in the projection domain. The optimization problem was solved using an alternating minimization scheme, where the image was updated through the alternating direction method of multipliers (ADMM), and the correcting vector was adjusted based on the updated image. Zhu et al.\cite{zhu2024dual} used a dual-domain regularization model to directly correct stripe artifacts in the projection domain by adjusting the compensation coefficients for detector response inconsistencies. They also applied smoothing constraints to the reconstructed image to further correct ring artifacts. However, these methods rely on forward and backward projection modeling, where inaccuracies in the forward model can introduce systematic biases, affecting the quality of the reconstructed image.

Deep learning-based methods for ring artifact removal offer new strategies from a data-driven perspective, leveraging large datasets with and without artifacts to learn how to remove ring artifacts. These methods have significant potential for development and promising application prospects. Chang et al.\cite{chang2020cnn} proposed a CNN-based method to reduce mixed ring artifacts in CT images by combining information from the image and sinogram domains. The CNN inputs the original and sinogram-corrected images and outputs an artifact-reduced image. Image mutual correlation is then used to fuse the sinogram correction with the CNN output, producing a hybrid corrected image. Liu et al.\cite{liu2023detector} introduced a low-dose CT ring artifact correction method based on detector shifting and deep learning. During CT scanning, the detector is randomly shifted horizontally at each projection to reduce ring artifacts in the pre-processing stage, converting the artifacts into dispersed noise. Deep learning is then used to denoise the dispersed and statistical noise, achieving ring artifact removal. Zhao et al.\cite{zhao2018removing} framed the ring artifact issue caused by detector pixel response inconsistencies as an ``adversarial problem", using a GAN-based image transformation strategy to remove ring artifacts from CT images through generative adversarial loss and custom smoothing loss. Wang et al.\cite{wang2019removing} employed GANs with unidirectional relative total variation loss, perceptual loss, and adversarial loss to effectively remove ring artifacts. However, the effectiveness of deep learning methods for artifact removal heavily relies on the size and quality of the training dataset. The lack of real ring artifact datasets and the discrepancy between simulated and real data lead to suboptimal artifact removal results.

\section{Method}
\begin{figure*}[!ht]
\centering
 \includegraphics[width=0.90\textwidth]{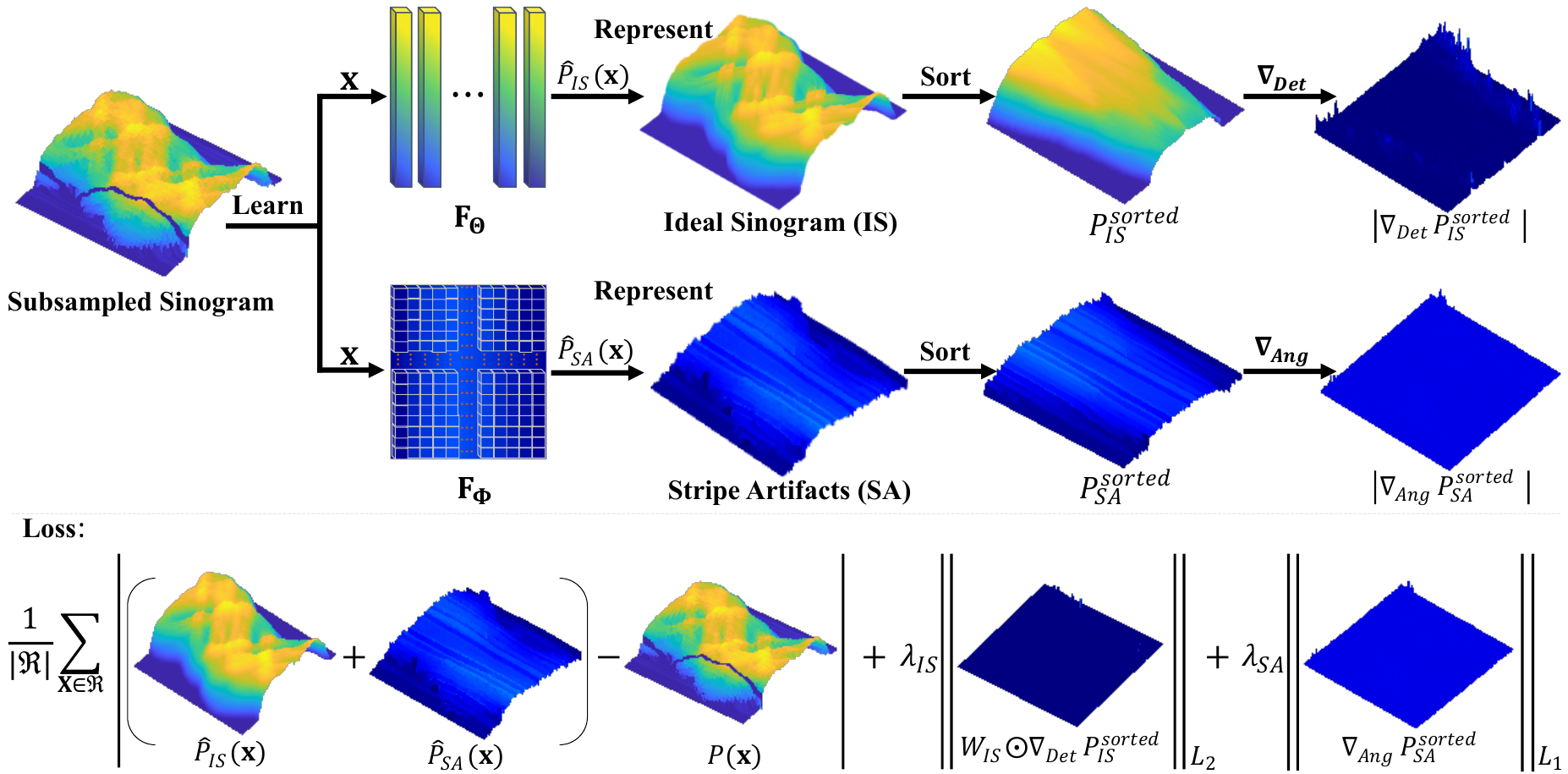}
\caption{Framework diagram for the implicit neural representation of sinogram data. The network $\mathbf{F}_\mathbf{\Theta}$ is used to parameterize the IS, while the network $\mathbf{F}_\mathbf{\Phi}$ represents the SA present in the sinogram. }
\label{fig_2}
\end{figure*}

This section will provide a detailed description of the proposed method. By utilizing INR to parameterize the sinogram, we decompose it into two parts: IS and SA. The proposed method framework for parameterizing the sinogram data is illustrated in Fig.~\ref{fig_2}.

In this section, defective and non-defective pixels are first labeled. Then, a forward model for the decomposition of IS and SA is established, incorporating effective regularization constraints for their separation. The design of the loss function is also presented, followed by a discussion on the residual compensation strategy and the network model selection. 

\subsection{Labeling Defective and Non-Defective Pixels}
To identify defective pixels, the gradient of the sinogram $P$ is computed in the angular direction, and its average value is compared to a predefined threshold $\mu=1\times10^{-6}$. Pixels with an average gradient value greater than $\mu$ are designated as non-defective (marked as $1$), while others are designated as defective (marked as $0$). The set of non-defective pixels is denoted by $\mathcal{R}$, and its indicator function $\mathbf{I}$ is defined as follows:
\begin{align}\label{Eq1}
\mathbf{I}_{i,j}=\left\{\begin{array}{cc} 1, & \overline{\nabla_{Ang} P}_{i,j} > \mu, \\
0, & \text {otherwise},
\end{array} \right.
\end{align}
where $i=1,2, \cdots, M$; $j=1,2, \cdots, N$; here, $\overline{\nabla_{Ang} P }_{i,j}$ is calculated as  $$\overline{\nabla_{Ang} P }_{i,j} = \frac{1}{N-1}\sum_{k=1}^{N-1} \left|P_{i,k}-P_{i,k+1}\right|,$$ with $M$ and $N$ representing the number of detector bins and view angles, respectively. 

\subsection{Implicit Neural Representation of Sinogram Data}
Sinogram is modeled as a 2-Dimensional (2D) continuous implicit function $P(\mathbf{x})$, which consists of IS $P_{IS}(\mathbf{x})$ and SA $P_{SA}(\mathbf{x})$. For $P_{IS}(\mathbf{x})$ and $P_{SA}(\mathbf{x})$, parameterization is performed using networks $\mathbf{F}_\mathbf{\Theta}$ and $\mathbf{F}_\mathbf{\Phi}$, respectively:
\begin{align}
\mathbf{F}_\mathbf{\Theta}: \mathbf{x}\mapsto {P_{IS}(\mathbf{x})},
\label{Eq2}
\end{align}
\begin{align}
\mathbf{F}_\mathbf{\Phi}: \mathbf{x}\mapsto {P_{SA}(\mathbf{x})},
\label{Eq3}
\end{align}
where $\mathbf{x}=(x, y)$ represents the 2D coordinates of the sinogram. The optimization model proposed in this work consists of two parts that are added together.
\begin{align}
P_{IS}(\mathbf{x}) + P_{SA}(\mathbf{x}) = P(\mathbf{x}).
\label{Eq4}
\end{align}

To ensure the effective parameterization of $P_{IS}(\mathbf{x})$ and $P_{SA}(\mathbf{x})$ by the networks $\mathbf{F}_\mathbf{\Theta}$ and $\mathbf{F}_\mathbf{\Phi}$, two unidirectional gradient regularization terms are introduced. A local smoothness term $\Psi_{IS}$ is applied along the detector direction for $P_{IS}$, while a sparsity constraint term $\Psi_{SA}$ is enforced along the angular direction for $P_{SA}$.

To prevent excessive smoothing at IS boundaries, the gradient of the sorted IS $P_{IS}^{sorted}$ in the detector direction is weighted by the normalized $P_{IS}^{sorted}$ values and constrained using an $\ell_2$ norm. The local smoothness term $\Psi_{IS}$ is defined as
\begin{align} \label{Eq5}
\Psi_{IS}(P_{IS})=\left\|W_{IS}\odot \nabla_{Det}P_{IS}^{sorted}\right\|_{2}, 
\end{align} 
where $\odot$ denotes element-wise multiplication. The gradient of $P_{IS}^{sorted}$ in the detector direction is
\[\resizebox{9cm}{!}{
$\nabla_{Det}{P_{IS}^{sorted}}_{i,j}=
\left\{\begin{array}{cc}
    {P^{sorted}_{IS}}_{i,j}-{P^{sorted}_{IS}}_{i+1, j},  & i=1,\cdots,M-1, \\
    {P^{sorted}_{IS}}_{i,j}-{P^{sorted}_{IS}}_{1, j}, & \quad i=M. 
\end{array}\right.$
}\]

The weighting coefficient is
$$W_{IS_{i,j}}= P_{IS_{i,j}}^{sorted}/\text{max}\left(P_{IS}^{sorted}\right).$$
The sparsity constraint term $\Psi_{IS}$ is given by
\begin{align} \label{Eq6}
\Psi_{SA}\left(P_{SA}\right)=\left\|\nabla_{Ang}P_{SA}^{sorted}\right\|_{1}, 
\end{align} 
where the gradient of sorted SA $P_{SA}^{sorted}$ in the angular direction is given by
\[\resizebox{9cm}{!}{
$\nabla_{Ang}{P_{SA}^{sorted}}_{i,j}=
\left\{\begin{array}{cc}
    {P^{sorted}_{SA}}_{i,j}-{P^{sorted}_{SA}}_{i, j+1},  & j=1,\cdots,N-1, \\
    {P^{sorted}_{SA}}_{i,j}-{P^{sorted}_{SA}}_{i, 1}, & \quad j=N. 
\end{array}\right.$
}\]

By incorporating the unidirectional regularization constraints $\Psi_{IS}$ and $\Psi_{SA}$, the separation of IS and SA within the sinogram is facilitated. Consequently, the loss function in this work comprises the Mean Absolute Error along with these two regularization constraints
\begin{align} \label{Eq7}
  \begin{aligned}
    \mathcal{L}& = \frac{1}{|\mathcal{R}|}\sum_{\mathbf{x} \in \mathcal{R}}\left|\left(\hat{P}_{IS}\left(\mathbf{x}\right)+\hat{P}_{SA}\left(\mathbf{x}\right)\right)-P\left(\mathbf{x}\right)\right|  \\ & + \lambda_{IS} \cdot \Psi_{IS}\left(\hat{P}_{IS}\right)+\lambda_{SA} \cdot  \Psi_{SA}\left(\hat{P}_{SA}\right), 
\end{aligned}
\end{align} 
where $|\mathcal{R}|$ is the number of elements in the dataset, $\hat{P}_{IS}$ and $\hat{P}_{SA}$ denote the predicted values for all pixels (both defective and non-defective), and $\lambda_{IS}$ and $\lambda_{SA}$ are the regularization coefficients. 

\subsection{Compensation of Residuals}
In addition to the components $\hat{P}_{IS}(\mathbf{x})$ and $\hat{P}_{SA}(\mathbf{x})$ predicted by the trained networks $\mathbf{F}_\mathbf{\Theta}$ and $\mathbf{F}_\mathbf{\Phi}$, respectively, there exists a small residual $E(\mathbf{x}) = P(\mathbf{x}) - \hat{P}_{IS}(\mathbf{x}) - \hat{P}_{SA}(\mathbf{x})$. Since $E(\mathbf{x})$ contains a small amount of structural information about the reconstructed object, it can be compensated into the output $P_{IS}^{out}$ to ensure image clarity. The compensation method is defined as
\begin{align} \label{Eq8} \begin{aligned} P_{IS}^{out}(\mathbf{x}) = \hat{P}_{IS}(\mathbf{x})+\kappa\cdot\hat{P}_{IS}(\mathbf{x})\cdot\tilde{E}(\mathbf{x}), \end{aligned} \end{align} 
where $\mathbf{x}$ represents the coordinates of non-defective pixels, $\kappa$ is the compensation coefficient, $\tilde{E}(\mathbf{x})$ is the residual information after subtracting its mean value in the angular direction.

Subtracting the mean in the angular direction removes most stripe noise from $E(\mathbf{x})$, leaving only the structural information of the object in $\tilde{E}(\mathbf{x})$. The amplitude of $\tilde{E}(\mathbf{x})$ is amplified by multiplying it with $\hat{P}_{IS}(\mathbf{x})$, enhancing the structural details in proportion to the projection values of $\hat{P}_{IS}(\mathbf{x})$, thereby improving image clarity. The parameter $\kappa$ adjusts the compensation intensity to enhance structure while minimizing noise and distortion.

\subsection{Network Model}
This paper requires two INR models to represent IS and SA separately. The core idea of INR \cite{2020NeRF, SIREN, NeFieldsinVisualComputingBeyond, essakine2024we, liu2024finer, kazerouni2024incode, saragadam2023wire, sun2021coil} is to train multi-layer perceptrons (MLPs) with activation functions to map input coordinates to the desired target values, thus learning continuous mapping functions from discrete data. However, while fully connected networks with positional encoding \cite{2020NeRF, sun2021coil} can record data information through weights, they are computationally inefficient, take a long time to train, and have limited capacity to express complex signals. Recent research, such as \cite{muller2022instant, chen2022tensorf, xie2023diner}, has explored the possibility of using structured storage of latent features and combining structured feature codes with lightweight MLPs networks, which improves the efficiency and expressiveness of INR.

\begin{figure}[!ht]
\centering
 \includegraphics[width=0.50\textwidth]{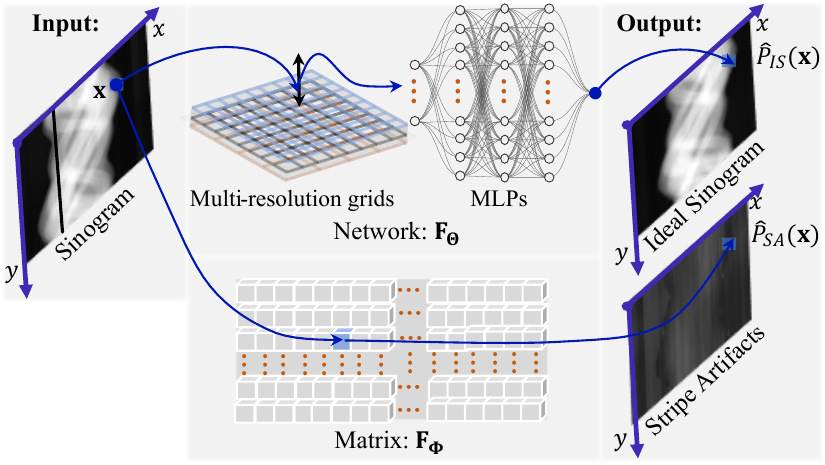}
\caption{Network model diagram. The network $\mathbf{F}_\mathbf{\Theta}$ composed of multi-resolution grids and MLPs is used to parameterize IS, while the learnable Matrix network $\mathbf{F}_\mathbf{\Phi}$ represents SA.}
\label{fig_3}
\end{figure}

For the network $\mathbf{F}_\mathbf{\Theta}$, the goal is to accurately reconstruct IS and predict the projection values of defective pixels. Given that IS typically exhibits low contrast and local smoothness, an ideal network should ensure continuity, effectively capture high-frequency details, converge rapidly, have low memory consumption, be insensitive to initialization, and remain stable during training. The priority for network $\mathbf{F}_\mathbf{\Phi}$ is its simplicity and fast convergence, without considering its continuous representation capability. Therefore, the simplest network model, a learnable parameter matrix, is chosen. Based on these requirements, we evaluated several representative network architectures \cite{2020NeRF, SIREN, fathony2020multiplicative, xie2023diner, muller2022instant}, and selected the one shown in Fig.~\ref{fig_3} by balancing convergence speed and representational accuracy.

The network model $\mathbf{F}_\mathbf{\Theta}$ consists of two components: a multi-resolution grid module and an MLP module. The grid module consists of three layers with resolutions of $\frac{M}{4} \times \frac{N}{4}$, $\frac{M}{3} \times \frac{N}{3}$, and $\frac{M}{2} \times \frac{N}{2}$, each vertex in these layers storing two feature vectors. The grid size is proportional to the number of detector bins $M$ and view angles $N$. The highest-resolution grid is set to half of $M \times N$, which enables effective modeling of defect regions while capturing high-frequency details. The number of grid layers and the dimension of feature vectors are selected based on a trade-off between representational capacity and computational cost. Although increasing these parameters may enhance reconstruction quality, it also leads to higher resource consumption. In this work, a three-level grid structure is adopted to achieve a favorable balance between model expressiveness and computational efficiency. The MLP module has three hidden layers, each with 64 neurons, and uses ReLU activation functions. While increasing the MLP's capacity (e.g., more layers or neurons) improves reconstruction quality, performance gains eventually plateau, with significant increases in training cost. A three-layer MLP with 64 neurons per layer achieves a good balance between accuracy and efficiency. Algorithm \ref{alg_1} details the complete process of the proposed method. 

\begin{algorithm}[!ht]
\caption{Training Details of the Proposed Method}
\label{alg_1}
\Input{Sinogram data $P$.}
\Output{Corrected Ideal Sinogram $P_{IS}^{out}$.}
Establish the coordinate system of $P$ and normalize its coordinates to the range $[-1, 1]$, obtaining the set $\mathcal{U}$\;
Extract the set of non-defective pixels $\mathcal{R}$ using Eq.~\eqref{Eq1}\;
Initialize parameters $\Theta$ for $\mathbf{F}_\mathbf{\Theta}$ and $\Phi$ for $\mathbf{F}_\mathbf{\Phi}$\;
\For{each training iteration}{
    Select a batch $\mathbf{x}_{b}$ from $\mathcal{U}$\;
    Compute $\hat{P}_{IS}(\mathbf{x}_{b})$ and $\hat{P}_{SA}(\mathbf{x}_{b})$ by applying $\mathbf{F}_\mathbf{\Theta}$ and $\mathbf{F}_\mathbf{\Phi}$\;
    Compute $\hat{P}_{IS}^{sorted}$ by sorting $\hat{P}_{IS}(\mathbf{x}_{b})$ along the angular axis and store the sorting index\;
    Compute $\hat{P}_{SA}^{sorted}$ by applying the sorting index from $\hat{P}_{IS}^{sorted}$ to $\hat{P}_{SA}(\mathbf{x}_{b})$\;
    Compute $\Psi_{IS}$ using Eq.~\eqref{Eq5} and $\Psi_{SA}$ using Eq.~\eqref{Eq6}\;
    Compute the total loss $\mathcal{L}$ using $P(\mathbf{x})$, $\hat{P}_{IS}(\mathbf{x})$, $\hat{P}_{SA}(\mathbf{x})$, $\Psi_{IS}$, and $\Psi_{SA}$ with Eq.~\eqref{Eq7}\;
    Update parameters: $(\Theta, \Phi) \leftarrow \mathrm{Adam}(\mathcal{L}, \Theta, \Phi)$;
}
Compute the final $\hat{P}_{IS}$ using $\Theta$ and Eq.~\eqref{Eq2}\;
Output the corrected $P_{IS}^{out}$ using Eq.~\eqref{Eq8}.
\end{algorithm}

\section{Experiments}
This section provides a comprehensive evaluation of the proposed method using both simulated and real data. Additionally, experiments are conducted to analyze the effects of different regularization terms and perform an ablation study on the INR representation.

To ensure a thorough and fair comparison, we do not include deep-learning-based methods due to the lack of publicly available benchmark datasets. Instead, we compare our method with several state-of-the-art model-based methods, including normalization-based \cite{Rivers2024}, regularization-based \cite{TITARENKO20101489}, FFT-based \cite{raven1998numerical,munch2009stripe}, and combined methods \cite{Vo:18}, as well as the methods proposed in \cite{an2020ring} and \cite{zhu2024dual}.

The PyTorch framework is used to implement the network models. The multi-resolution grid module of networks $\mathbf{F}_\mathbf{\Theta}$ and $\mathbf{F}_\mathbf{\Phi}$ is initialized with uniform distributions in the range of $\pm 1 \times 10^{-4}$ to promote training stability. The Adam optimizer \cite{kingma2014adam} is used during training, with an initial learning rate of $1 \times 10^{-4}$ for both networks.  The regularization coefficients $\lambda_{IS}$ and $\lambda_{SA}$ start at $1 \times 10^{-4}$ and are linearly increased to $5 \times 10^{-3}$ and $1 \times 10^{-3}$, respectively, as training progresses. In the early stages, large regularization values may cause the projection values of IS and SA to drop to zero, hindering learning. As training advances and the projection values increase, the regularization coefficients are adjusted to maintain stable learning. During the experiments, the sinogram is normalized before training, and after training, the predicted data from the INR undergoes inverse normalization. The parameter $\kappa$ is set to 1, and training is performed for 5,000 iterations. All experiments were conducted on an NVIDIA GeForce RTX 3080 GPU with 10GB of memory.

\subsection{Experiments with Numerical Simulation}

\begin{figure}[!ht]
\centering
 \includegraphics[width=0.40\textwidth]{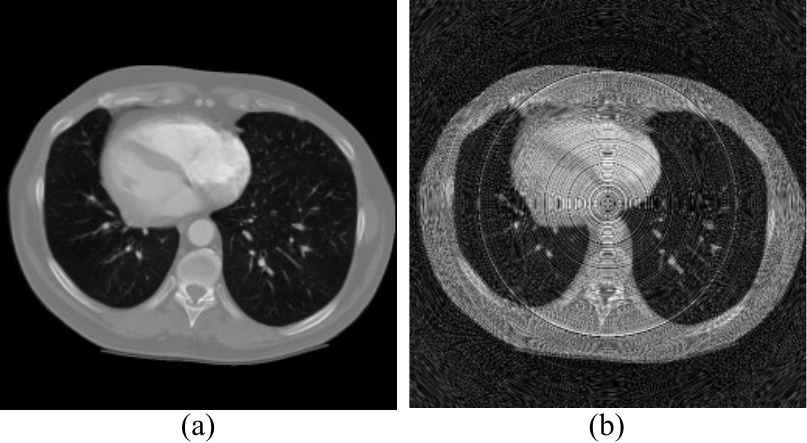}
\caption{Simulated phantom. (a) Phantom. (b) Result of FBP from Fig.~\ref{fig_5} (b). The display range is [0, 0.02].}
\label{fig_4}
\end{figure}

\begin{figure}[!ht]
\centering
 \includegraphics[width=0.45\textwidth]{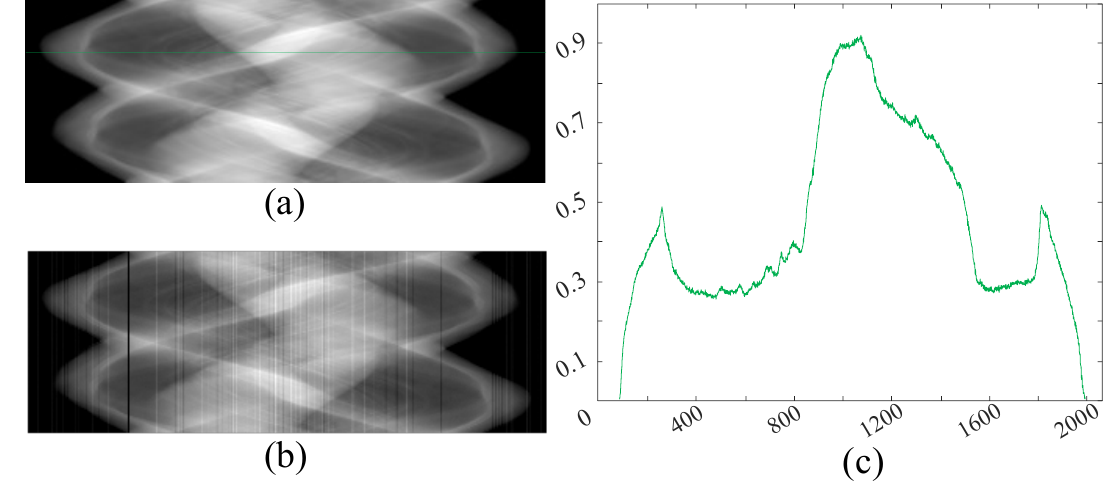}
\caption{Simulated sinogram. (a) Sinogram; (b) Sinogram with artifacts; (c) Profile curve of the 200th row in (a). The display range is [0, 1].}
\label{fig_5}
\end{figure}

The phantom in the simulation data is an abdominal CT slice\cite{mccollough2016tu}. In the scanning configuration, the distance between the X-ray source and the detector is 416.696 mm. The detector consists of 2068 bins, each with a length of 0.075 mm. The distance between the X-ray source and the rotation center is 297.143 mm. Uniform sampling is performed at 720 view angles within the $[0, 2\pi]$ range. The sinogram was generated by simulating fan-beam projection \cite{Biguri_2016}, as shown in Fig.~\ref{fig_5} (a). According to Eq.~\eqref{Eq1_}, $\pm10\%$ photon fluctuations were added to half of the randomly selected pixels to simulate stripe artifacts, and the projection values in columns 400 to 404 were set to zero to simulate defective pixels. The initial photon intensity is set to $1\times10^5$, and Poisson noise with the corresponding intensity is then added in the photon domain, as shown in Fig.~\ref{fig_5} (b). The result of FBP reconstruction using the sinogram with artifacts is displayed in Fig.~\ref{fig_4} (b). Fig.~\ref{fig_5} (c) shows the profile curve of the 200th row in Fig.~\ref{fig_5} (a). 
For the simulation data experiments, the performance of the methods is quantitatively assessed using the Peak Signal-to-Noise Ratio (PSNR) and Structural Similarity Index (SSIM)\cite{wang2004image}. 

\begin{figure*}[!ht]
\centering
 \includegraphics[width=0.80\textwidth]{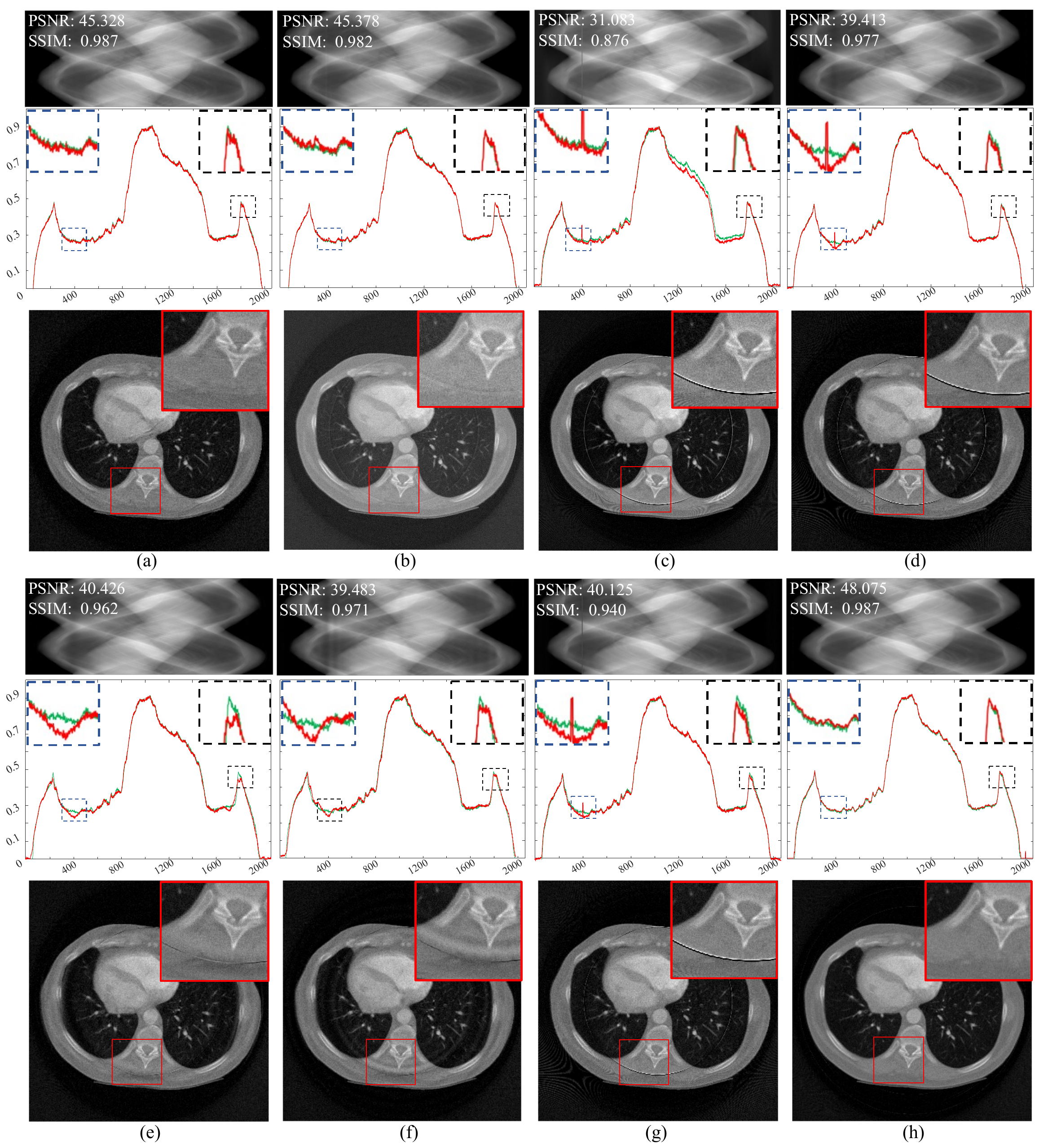}
\caption{Comparison experiments with simulated data. (a) Combination method. (b) Method from \cite{an2020ring}. (c) Method from \cite{zhu2024dual}. (d) Normalization method. (e) FFT method. (f) W-FFT method. (g) Regularization method. (h) Proposed method. The display range is [0, 1] for the sinogram and [0, 0.02] for the CT images.}
\label{fig_6}
\end{figure*}

The experimental results are shown in Fig.~\ref{fig_6}, which presents the corrected sinograms and provides a quantitative comparison with the reference sinogram regarding PSNR and SSIM. Higher scores indicate better consistency with the reference sinogram. The second row displays the profile curves of the 200th row, with the green curve representing the reference profile and the red curve showing the predicted profile. The third row shows the reconstructed images and their local zoom-in views.

Zoom-in views (b) to (g) in Fig.~\ref{fig_6} reveal some ring artifacts, indicating limitations in suppressing defective pixels with these methods. In contrast, the zoom-in views in Figs.~\ref{fig_6} (a) and (h) show reduced ring artifacts, with Fig.~\ref{fig_6} (h) exhibiting the least. The image in Fig.~\ref{fig_6} (a) presents other types of artifacts. In the profile curves of the 200th row, Figs.~\ref{fig_6} (c) to (g) show noticeable deviations from the reference profile curve in columns 400 to 404. Quantitative evaluation results show that the PSNR value of the predicted IS in Fig.~\ref{fig_6} (h) is the highest among all compared methods. The SSIM value of Fig.~\ref{fig_6} (h) is the same as that of Fig.~\ref{fig_6} (a), both being the highest, indicating that Fig.~\ref{fig_6} (h) demonstrates superior performance in signal fidelity and stripe artifact suppression.

\begin{figure*}[!ht] 
    \centering \includegraphics[width=0.8\textwidth]{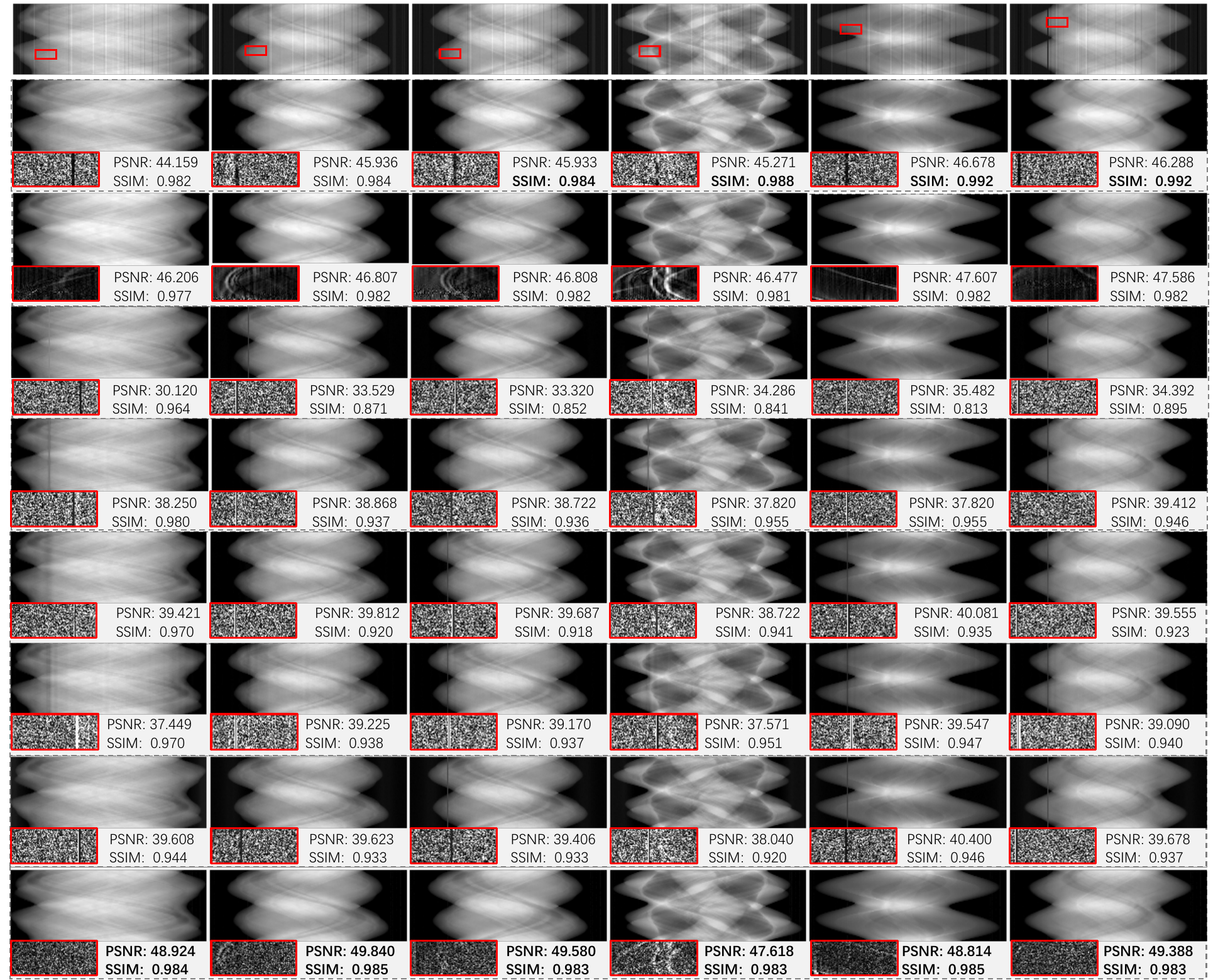} \caption{Supplementary simulation experiments. Lines 2–9 correspond to: the combination method, the method from \cite{an2020ring}, the method from \cite{zhu2024dual}, the normalization method, the FFT method, the W-FFT method, the regularization method, and the proposed method. The display range is [0, 1] for the sinogram and [0, 0.01] for the zoomed-in view of the local gradient.} \label{fig_R1_7} 
\end{figure*}

\begin{table}[h]
\centering
\caption{Average PSNR and SSIM metrics for the compared methods in Fig. \ref{fig_R1_7}. Methods 1-8 correspond to the second to ninth rows in Fig. \ref{fig_R1_7}.}
\label{RA_table_3_1}
\renewcommand{\arraystretch}{1.25}
\begin{tabular}{c@{\hspace{2pt}}c@{\hspace{2.3pt}}c@{\hspace{2.3pt}}c@{\hspace{2.3pt}}c@{\hspace{2.3pt}}c@{\hspace{2.3pt}}c@{\hspace{2.3pt}}c@{\hspace{2.3pt}}c}
\hline\hline 
Method & 1 & 2 & 3 & 4 & 5 & 6 & 7 & 8 \\
\hline PSNR & 45.711 & 46.915 & 33.522 & 38.482 & 39.546 & 38.675 & 39.459 & \textbf{49.027} \\
\hline SSIM & \textbf{0.987} & 0.981 & 0.873 & 0.952 & 0.935 & 0.947 & 0.936 & \textbf{0.984} \\
\hline\hline 
\end{tabular}
\end{table}
    
 We randomly selected six additional phantoms from the low-dose dataset \cite{mccollough2016tu} and simulated stripe artifacts to further evaluate the performance of different methods in enhancing image quality. Fig.~\ref{fig_R1_7} presents the sinograms with stripe artifacts and the results after processing by different methods, along with the PSNR and SSIM metrics for each method. Additionally, we provide gradient maps along the detector direction for the zoomed-in regions (red box). By comparing the PSNR and SSIM values, it can be observed that our proposed method consistently achieves higher scores in most cases. Furthermore, the zoomed-in gradient maps show the least stripe artifacts for our method, demonstrating its superior performance in artifact removal and image detail restoration.

 Additionally, we provide the average PSNR and SSIM metrics for the various comparison methods based on the experimental data in Fig.~\ref{fig_R1_7}, as shown in Table \ref{RA_table_3_1}. These results demonstrate that our method consistently exhibits strong performance in most cases. The relatively weaker SSIM values can be attributed to the weighting scheme of $W_{IS}$, which assigns lower weights to regions with low projection values near the sinogram boundaries. These regions contain little meaningful projection information of the reconstructed object, thus contributing minimally to the final CT image quality. However, this weighting scheme may reduce the smoothness constraint on $\Psi_{IS}$ in these areas, which affects the SSIM values to a certain extent, although it does not significantly impact the overall CT image reconstruction quality.

In summary, the proposed method performs well in suppressing stripe artifacts and accurately estimating the IS values of defective pixels, thereby improving the quality of image reconstruction to a certain extent.

\subsection{Experiments with Real Data}

\begin{figure}[!ht]
\centering
\includegraphics[width=0.48\textwidth]{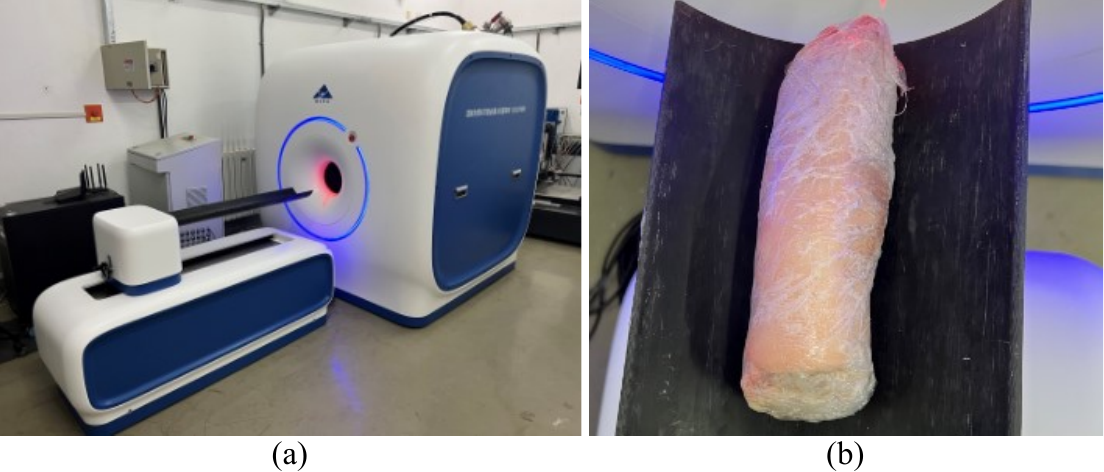}
\caption{The experimental PCD-CT system platform and the scanned sample. (a) PCD-CT system platform. (b) Scanned fresh pork trotter sample.}
\label{fig_13}
\end{figure}

\begin{figure}[!ht]
\centering
\includegraphics[width=0.40\textwidth]{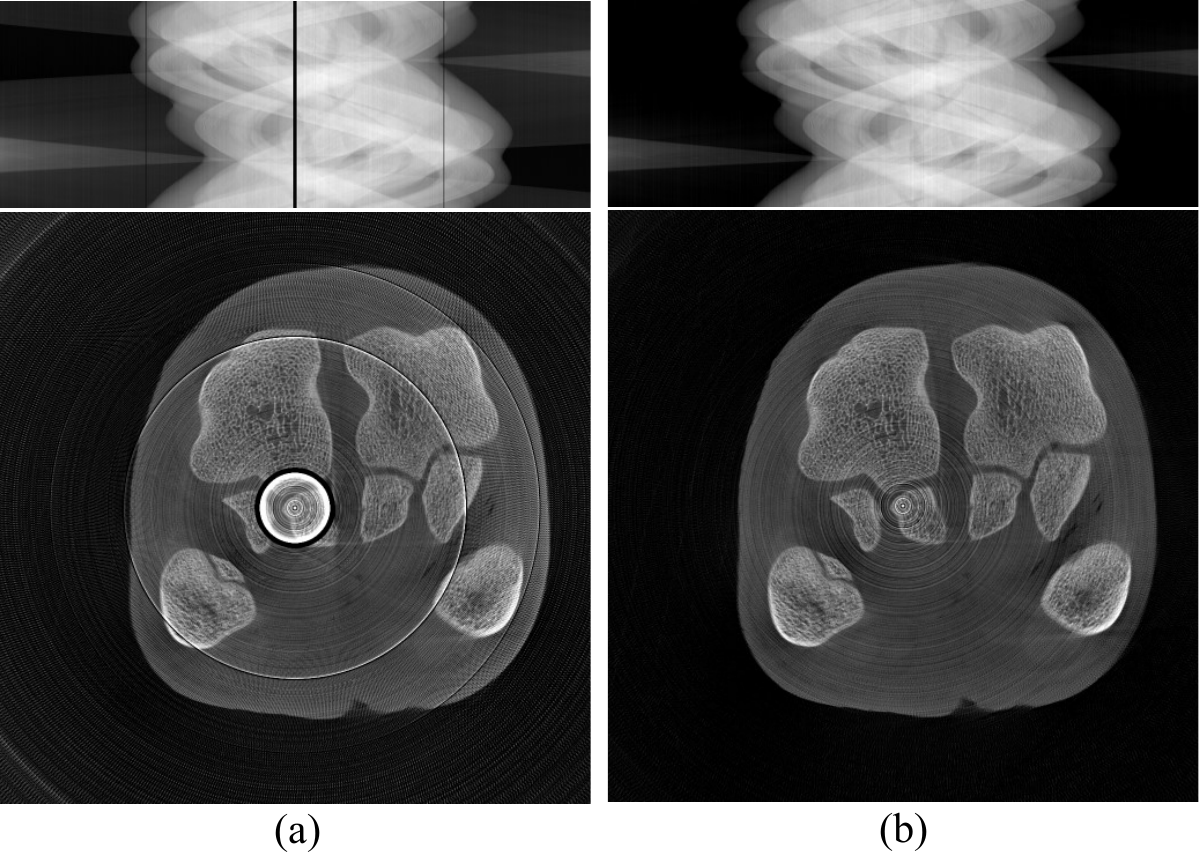}
\caption{Real data and FBP results. (a) Unprocessed sinogram and its CT image. (b) Sinogram with black gaps removed using linear interpolation and its CT image. The display range is [0, 3.144] for the sinograms and [0, 0.1] for the CT images.}
\label{fig_7}
\end{figure}

\begin{figure*}[!ht]
\centering \includegraphics[width=0.80\textwidth]{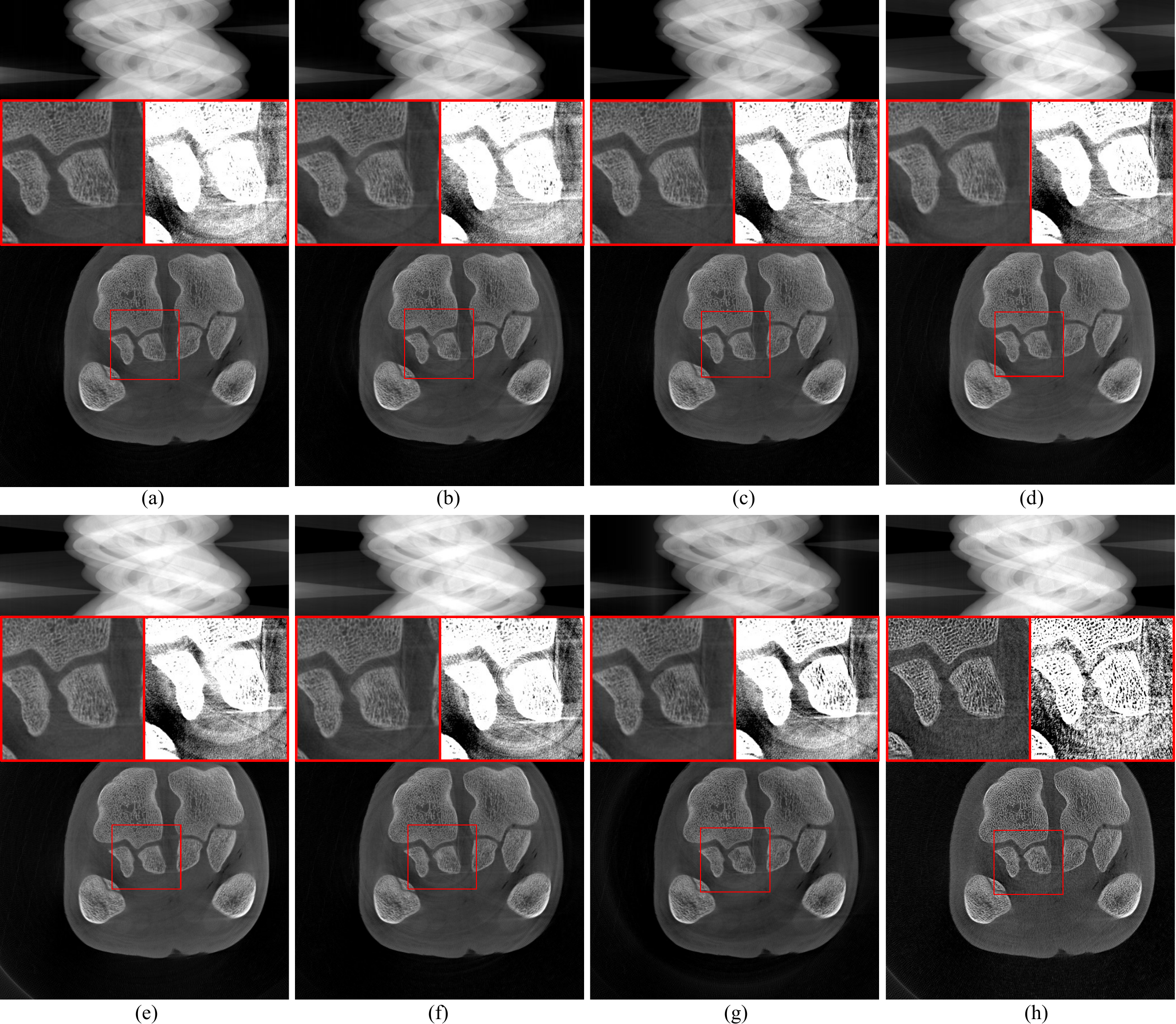}
\caption{Comparison experiments with real data. (a) Combination method. (b) Method from \cite{an2020ring}. (c) Method from \cite{zhu2024dual}. (d) Normalization method. (e) FFT method. (f) W-FFT method. (g) Regularization method. (h) Proposed method. The display range is [0, 3.144] for the sinograms and [0, 0.10] for the CT images. The zoom-in view on the left shows [0, 0.1], and on the right, it shows [0.016, 0.030].}
\label{fig_8}
\end{figure*}

The real data experiment was conducted using a 120 kV X-ray source (Libra13UlNE, iRay, China) to generate X-rays. The detector used was a Flat-Panel Photon Counting Detector (EIGER2, DECTRIS, Switzerland). The scanned sample was a fresh pork trotter. Fig.~\ref{fig_13} shows the PCD-CT system and the scanned sample. The scanning configuration was consistent with that of the simulated data experiments. Fig.~\ref{fig_7} shows the sinogram data collected during the experiment. The first row of Fig.~\ref{fig_7} (a) shows the reference sinogram, where black stripes caused by the gaps between the FPCD panels can be observed. Most existing stripe artifact removal methods struggle to handle artifacts caused by defective pixels, such as these black gaps. To comprehensively evaluate the performance of the proposed method, linear interpolation was used to preprocess these black gaps. The first row of Fig.~\ref{fig_7} (b) shows the preprocessed sinogram. To ensure fairness in the comparison experiments, the relevant parameters of the compared methods were fine-tuned from their default settings to achieve optimal stripe artifact removal performance. 

Figs.~\ref{fig_8} (a), (b), and (h) show the results of stripe artifact removal from the sinogram containing black gaps in Fig.~\ref{fig_7} (a). All methods successfully removed the black stripes and reduced other stripe artifacts. The results of the combination method \cite{Vo:18} in Fig.~\ref{fig_8} (a) demonstrate its effectiveness in removing stripe artifacts. The overall image is clear, and details are well-preserved. However, in the local zoomed-in regions, some ring artifacts remain. Fig.~\ref{fig_8} (b) shows the result after applying flat-field correction and linear interpolation for defective pixels (black gaps), followed by TV-$\ell_1$ aG filtering. The results indicate that while details are well preserved, noticeable ring artifacts remain in the local zoomed-in regions. Fig.~\ref{fig_8} (h) presents the results of the proposed method, where defective pixels (those at the black gaps) were not involved in the neural network training. These defective pixel values were predicted by the network $\mathbf{F}_\mathbf{\Theta}$. The proposed method can maintain image continuity while filling gaps and effectively controlling ring artifacts. Figs.~\ref{fig_8} (c) to (g) show the experiments on stripe artifact removal from the sinogram after linear interpolation of the black gaps in Fig.~\ref{fig_7} (b). Fig.~\ref{fig_8} (c) uses a dual-domain regularization iterative method. The reconstructed results still show slight ring artifacts in the local zoomed-in regions. Although Figs.~\ref{fig_8} (c) to (g) methods improved the quality of the reconstructed images and reduced ring artifacts to some extent, residual ring artifacts can still be observed in the local zoomed-in regions. The proposed method in Fig.~\ref{fig_8} (h) best preserves image continuity and reduces ring artifacts.

In summary, when comparing the methods in Fig.~\ref{fig_8}, the method in Fig.~\ref{fig_8} (a) is relatively complex, achieving good stripe artifact removal results but still leaving slight ring artifacts. The methods in Figs.~\ref{fig_8} (b) and  (c) require extensive parameter tuning to achieve satisfactory results. Methods in Figs.~\ref{fig_8} (d) to  (g) leave some residual ring artifacts. Fig.~\ref{fig_8} (h) provides the best ring artifact removal and the clearest reconstructed image. Comparative analysis shows that the proposed method effectively removes ring artifacts, whereas other methods leave varying degrees of ring traces in the reconstructed images. 

\subsection{Verification of Regularization Terms}
\begin{figure*}[!ht]
\centering
\includegraphics[width=0.80\textwidth]{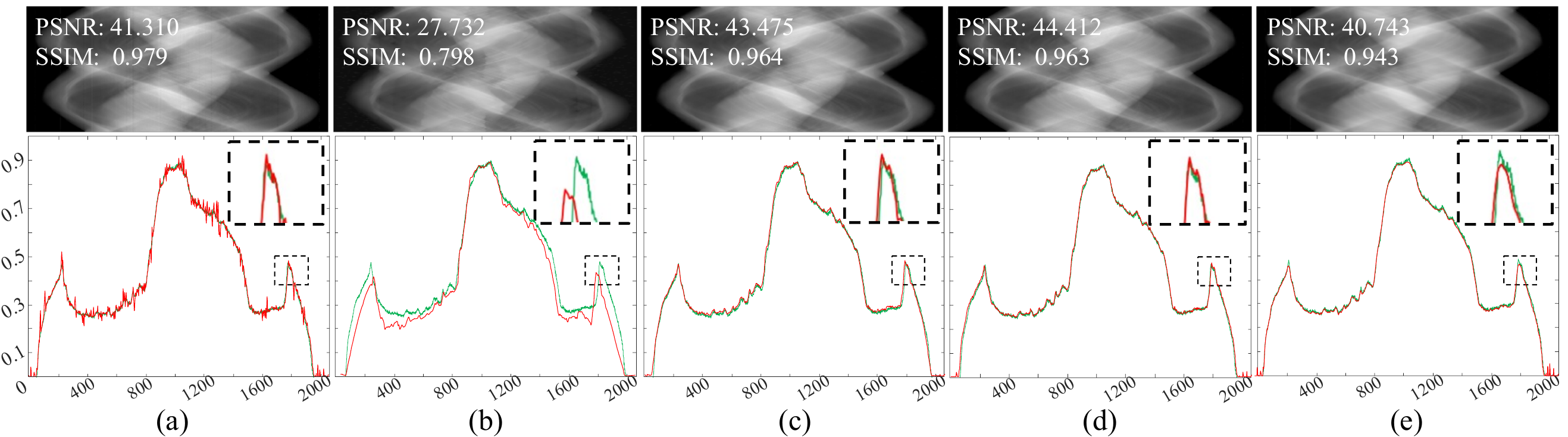}
\caption{Impact of unidirectional gradient constraints on the INR-parameterized IS: (a) Angular gradient with $\ell_1$ norm constraint on SA only;
(b) Detector gradient with $\ell_2$ norm constraint on IS only;
(c) Angular $\ell_1$ norm constraint on SA combined with detector $\ell_2$ norm constraint on IS;
(d) Same as (c), but with a weighted detector 
$\ell_2$ norm constraint;
(e) Unsorted SA and IS, with combined angular 
$\ell_1$ norm constraint and detector $\ell_2$ norm constraint. The disparity range is [0, 1].}
\label{fig_9}
\end{figure*}

To verify the effectiveness of the two unidirectional gradient constraints and their role in removing ring artifacts from CT images through the sorted sinogram, we used the simulated sinogram shown in Fig.~\ref{fig_5} (b). Fig.~\ref{fig_9} presents the results obtained by applying different unidirectional gradient constraints to INR, with the number of training iterations fixed at 10,000. The first rows of Fig.~\ref{fig_9} show the INR-parameterized IS, respectively, with PSNR and SSIM scores annotated in the figures. The second rows display the profile lines of the 200th row. The green lines represent the reference profile, while the red lines represent the profile lines of the INR-parameterized IS.

From the observations in Figs.~\ref{fig_9} (a) and  (b), it is clear that the IS shown in Fig.~\ref{fig_9} (a) exhibits significant stripe artifacts. However, the IS of Fig.~\ref{fig_9} (b) shows less prominent stripe artifacts, although the profile plot reveals that the IS profile is overly smooth. This indicates that applying only the angular direction gradient $\ell_1$ norm constraint on the SA is insufficient for effectively removing stripe artifacts. Adding a gradient $\ell_2$ norm constraint in the detector direction on the IS can somewhat reduce stripe artifacts, but may also lead to overly smooth images. Further observation of Figs.~\ref{fig_9} (c) and (d) shows that all IS effectively mitigate stripe artifacts. However, the quantitative metrics indicate that the PSNR and SSIM of Fig.~\ref{fig_9} (d) are higher than those of Fig.~\ref{fig_9} (c), demonstrating the effectiveness of the detector-direction gradient weighting strategy.

In contrast, the profile lines of Fig.~\ref{fig_9} (e) in columns 200 and 1800 do not effectively match the reference profile. In comparison, the IS profile of Fig.~\ref{fig_9} (d) more closely aligns with the reference profile. This suggests that applying a weighted gradient $\ell_2$ norm constraint in the detector direction on the sorted IS and an angular gradient $\ell_1$ norm constraint on the sorted SA effectively removes stripe artifacts while maintaining image edge sharpness. Comparing Fig.~\ref{fig_6} (h) and Fig.~\ref{fig_9} (d) reveals that the use of the residual compensation strategy further enhances the PSNR.

\subsection{Ablation Study on the Effectiveness of INR}
\begin{figure}[!ht]
\centering
\includegraphics[width=0.48\textwidth]{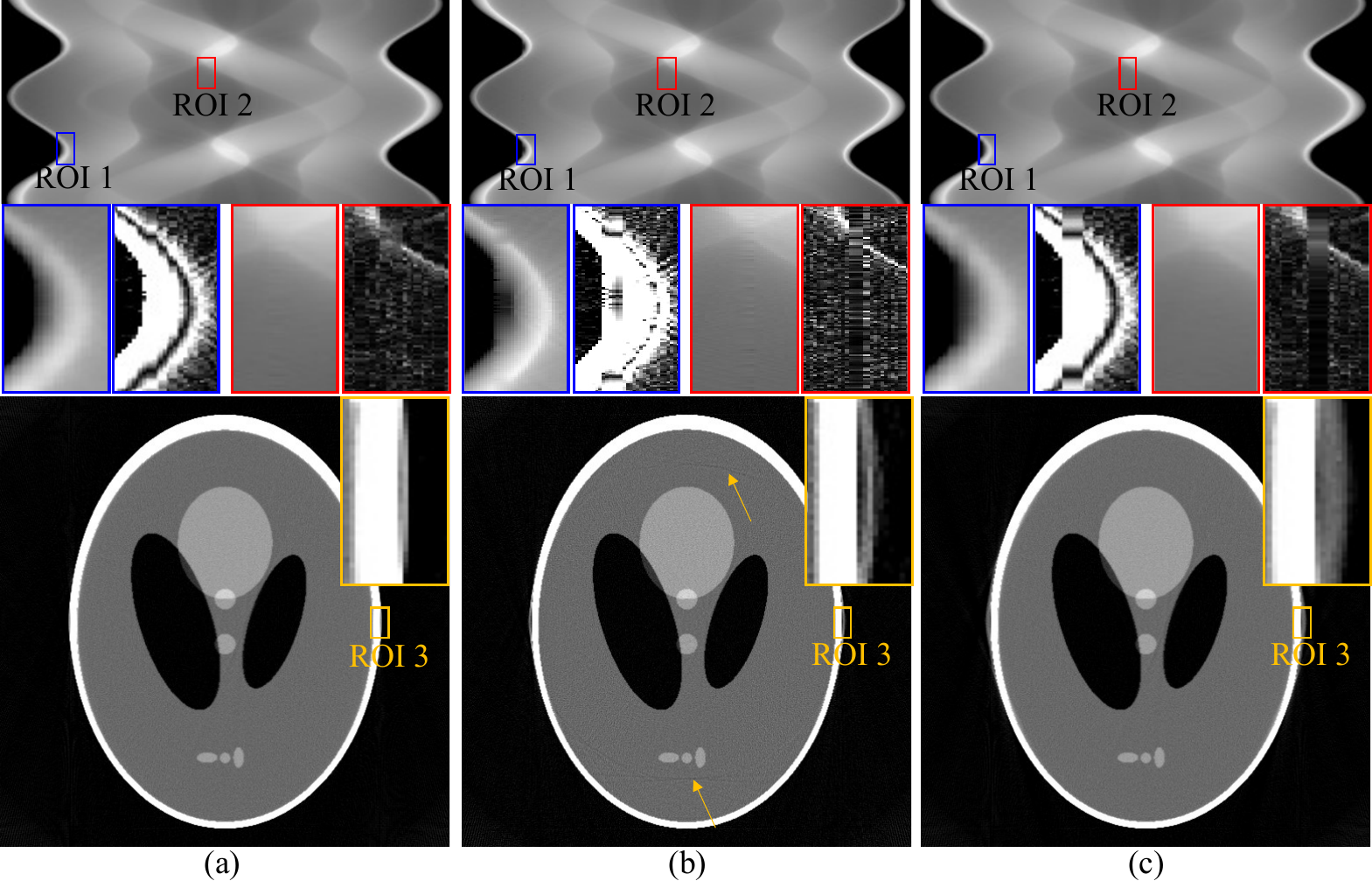}
\caption{Comparison of different representation methods: (a) INR with multi-resolution grids and MLPs; (b) learnable matrix; and (c) linear interpolation. The first row shows sinograms (display range: [0, 1]); the second row shows zoomed-in views of ROI 1 and ROI 2 (display range: [0, 1]), and their corresponding gradient maps (display range: [0, 0.015]); the third row presents reconstructed CT images and a zoomed-in view of ROI 3 (display range: [0, 0.015]).}
\label{fig_R1}
\end{figure}

To further evaluate the effectiveness of the INR method, we compare three approaches for handling unresponsive stripes: INR, a learnable matrix (analogous to the SA component), and linear interpolation. In the simulation, two 10-pixel-wide unresponsive stripes were introduced at the center and boundary of the sinogram, based on the characteristics observed in real-world CT systems. The corresponding results are presented in Fig.~\ref{fig_R1}.

ROI 1 is located at the edge of the sinogram, where the projection values vary considerably. From the zoomed-in view and its gradient map along the detector direction, both the learnable matrix (b) and linear interpolation (c) yield noticeable errors, with the latter exhibiting the most severe degradation. In contrast, the INR-based result (a) better preserves structural details and yields more accurate estimations. ROI 2 lies at the center of the sinogram, where projection values vary more smoothly. The zoomed-in views of this region reveal minor visual differences among the three methods. However, the corresponding gradient maps indicate that linear interpolation introduces distinct stripe artifacts, while the learnable matrix provides moderate improvement. Again, the INR method achieves the smoothest and most consistent result. The third row shows reconstructed CT images, where ROI 3 corresponds to the projection region of ROI 1. Linear interpolation results in rough and inaccurate boundaries (c), and the learnable matrix performs slightly better (b). The INR method (a), however, delivers the highest reconstruction quality.

INR consistently outperforms the other methods in mitigating unresponsive stripe artifacts. While explicit regularization imposes useful constraints, the implicit smoothness of INR further enhances estimation continuity and reduces artifacts. This advantage is especially evident in regions near the sinogram boundary.

\section{Discussions}
The experimental results demonstrate that the proposed method performs better than existing methods regarding image clarity, continuity, and ring artifact removal. This section discusses the strengths and limitations of the proposed method.

Regarding image clarity, as shown in Fig.~\ref{fig_8}, the proposed method produces the clearest reconstructed images. This can be attributed to the effective separation of IS and SA, as well as the compensation of residuals. The residual compensation plays a role in enhancing image details and contrast.  However, as shown in Fig.~\ref{fig_8} (h), the compensation process may also amplify noise, primarily due to the presence of scatter noise in real data, which was not explicitly modeled. Regarding image continuity, as illustrated in Fig.~\ref{fig_6}, the proposed method successfully predicts the IS values for gaps caused by non-responsive black stripes using the continuous representation capabilities of the implicit neural network. As a result, no residual rings caused by these gaps are visible in the reconstructed images. Both simulation and real data experiments show that the proposed method results in the least residual ring artifacts.

Despite its excellent performance, the proposed method has some limitations. While effective in separating SA from IS, the unidirectional gradient regularization framework may lead to information loss, including minor stripe artifacts and small structural details of the object. A compensation scheme is proposed in Eq.~\eqref{Eq8} to address this limitation. While this scheme is simple and effective in enhancing image clarity and contrast, it does introduce some noise. Future research could explore more sophisticated iterative schemes to progressively incorporate residual information into the SA. Additionally, for PCD data, it will be important to consider modeling the fixed pattern noise.

\section{Conclusion}
This study proposed a method for effectively separating stripe artifacts from the sinogram using INR parameterization. Applying unidirectional gradient regularization to the SA and IS effectively separated the two. The $\ell_1$ norm regularization of the gradient in the angular dimension enhances the stripe artifact features, while the $\ell_2$ norm regularization of the weighted gradient in the detector direction preserves the local smoothness of the data. Additionally, by utilizing the continuous representation capability of INR, the proposed method predicts defective pixel projection values using only the inconsistent detector response to X-ray, without requiring additional input. The experimental results demonstrate the effectiveness of the technique in removing ring artifacts from CT images. As a novel ring artifact removal technique, this INR parameterization strategy successfully demonstrates its effectiveness in enhancing CT image quality, providing a new perspective for addressing the challenge of ring artifacts.

\bibliography{ref.bib}
\bibliographystyle{IEEEtran}

\end{document}